\documentclass[acmsmall,screen]{acmart}
\usepackage[linesnumbered,ruled,lined]{algorithm2e}
\usepackage{booktabs}
\usepackage{caption}
\usepackage{subcaption}
\usepackage[inkscapelatex=false]{svg}
\usepackage{multirow}
\usepackage{amsmath}
\usepackage{pifont}
\usepackage{array}
\usepackage{makecell}
\usepackage[table]{xcolor}
\usepackage{setspace}
\usepackage{enumitem}
\usepackage{tabularx}
\usepackage{microtype}
\usepackage[most]{tcolorbox}
\usepackage{threeparttable}
\usepackage{graphicx} 
\usepackage[export]{adjustbox} 
\usepackage{listings}
\usepackage{rotating} 
\newcolumntype{Y}{>{\raggedright\arraybackslash}X} 
\usepackage{amsmath}  
\usepackage{pifont}  
\usepackage{color}
\usepackage{tikz}
\usepackage{hyperref} 
\usepackage{cleveref} 
\usetikzlibrary{trees}

\definecolor{mygray}{RGB}{128,128,128}
\definecolor{darkred}{rgb}{0.8, 0.1, 0.1}
\definecolor{darkgreen}{rgb}{0.1, 0.6, 0.1}
\definecolor{lightgray}{gray}{0.9}
\definecolor{darkgray2}{RGB}{90,90,90}
\definecolor{codegreen}{RGB}{56,157,60}
\definecolor{codeblue}{RGB}{19,119,199}
\definecolor{codepurple}{RGB}{162,65,163}
\definecolor{codeorg}{RGB}{240,136,26}
\definecolor{codegreen2}{RGB}{0,193,6}

\hypersetup{colorlinks = true, 
	    linkcolor = blue, 
	    urlcolor = blue,
            citecolor = red} 
\restylefloat{figure}
\AtBeginDocument{%
  }

\setcopyright{none}
\copyrightyear{2025}
\acmYear{2025}
\acmDOI{XXXXXXX.XXXXXXX}
\acmConference[Conference FSE '26]{The ACM International Conference on the Foundations of Software Engineering }{July 05--09,
  2026}{Montreal, Canada}
\acmISBN{978-1-4503-XXXX-X/2025/08}

\begin{document}

\title{Beyond Language Boundaries: Uncovering Programming Language Families for Code Language Models}

\author{Shangbo Yun}
\email{sjtu_yun@sjtu.edu.cn}
\affiliation{
\institution{School of Computer Science, Shanghai Jiao Tong University}
\country{China}
}
\author{Xiaodong Gu}
\email{xiaodong.gu@sjtu.edu.cn}
\affiliation{
\institution{School of Computer Science, Shanghai Jiao Tong University}
\country{China}
}
\author{Jianghong Huang}
\email{jh.huang@sjtu.edu.cn}
\affiliation{
\institution{School of Computer Science, Shanghai Jiao Tong University}
\country{China}
}
\author{Beijun Shen}
\authornote{Corresponding author.}
\email{bjshen@sjtu.edu.cn}
\affiliation{%
  \institution{School of Computer Science, Shanghai Jiao Tong University}
  \country{China}
}

\newcommand{\todoc}[2]{{\textcolor{#1}{\textbf{[#2]}}}}
\newcommand{\todoblue}[1]{\todoc{blue}{\textbf{#1}}}
\newcommand{\todored}[1]{\todoc{red}{\textbf{#1}}}
\newcommand{\todoorange}[1]{\todoc{orange}{\textbf{#1}}}

\newcommand{\yun}[1]{\todoblue{yun: #1}}
\newcommand{\huang}[1]{\todoblue{huang: #1}}
\newcommand{\new}[1]{{#1}}
\newcommand{\gu}[1]{\todored{gu: #1}}
\newcommand{\shen}[1]{\todored{shen: #1}}

\begin{abstract}
The rapid proliferation of diverse programming languages presents both opportunities and challenges for developing multilingual code LLMs. 
While existing techniques often train code LLMs by simply aggregating multilingual code data, few explore the deeper relationships between programming languages and how such relationships can be utilized to optimize the training and inference of code LLMs.  
In this work, we investigate two fundamental questions: (1) What are the deep linguistic relationships among programming languages? and (2) How can these relationships be leveraged to improve multilingual code LLMs? We propose an embedding-based framework to uncover the latent families of programming languages. 
Our approach begins by defining 21 primary linguistic features of programming languages, such as variable definition, control structures, and method declarations, and then employs LLMs to generate feature-aligned code samples across multiple languages. By embedding these semantically parallel code snippets from 19 languages, we construct a similarity matrix and perform hierarchical clustering to uncover inherent language relationships. Our analysis reveals clear hierarchical structures among programming languages. Closely related languages form well-defined clusters (\emph{e.g.}, C, C++, Java, and Swift group together), while Go exhibits as a central language with the highest cross-language similarity.
Building on the uncovered language families, we propose three strategies to enhance multilingual LLM training: transfer learning across linguistically related languages, linguistic proximity-guided curriculum learning, and centroid-based intermediary code translation. 
Experiments on four code intelligence tasks demonstrate that our methods significantly improve multilingual LLM performance. This work offers a universal perspective on programming languages and advances more effective strategies for multilingual code LLM training.

\end{abstract}


\maketitle

\section{Introduction}

In recent years, the landscape of programming languages has expanded dramatically, with an increasing number of languages emerging to address diverse computational challenges \cite{pragmatics}. Each language brings unique strengths tailored to specific domains, paradigms, and use cases. From the low-level efficiency of C language to the high-level abstractions of Python, and from the functional elegance of Haskell to the concurrency prowess of Go, the diversity of programming languages reflects the multifaceted nature of modern software development. However, this proliferation also raises critical questions about the relationships between languages, the criteria for language selection~\cite{KatzyID23}, and the potential for effectively integrating multiple languages~\cite{guo2022unixcoder}. Understanding these aspects is crucial not only for developers but also in the context of large language models (LLMs), which are reshaping the way we interact with and reason about code.

While existing approaches to training code LLMs largely rely on the simple integration of multilingual code data \cite{roziere2023code, deepseekcoder2024, li2023starcoder, ahmed2022multilingual}, they rarely explore the deeper linguistic relationships among programming languages or how these relationships could be leveraged to enhance model training. These relationships span multiple aspects such as syntactic structures and programming paradigms. For example, languages in the C-family (like C, C++, Java, and C\#) share highly similar control flow constructs, while functional languages such as Haskell and Scala emphasize immutability and higher-order functions.
Although such relationships may have been implicitly leveraged in prior multilingual LLM training, they have not yet been systematically modeled or explicitly studied. 
These gaps in current research limit our ability to harness the multilingual potential of code LLMs fully.

In this paper, we seek to answer two research questions:  
\textbf{RQ1} \textit{What are the deeper linguistic relationships between programming languages?} 
 and \textbf{RQ2} \textit{How can these linguistic relationships be leveraged to enhance the training and inference of code LLMs?} 

Answering these questions has significant implications for LLM training and the future design of programming languages. For instance, how do languages within the same family (\emph{e.g.}, C-based languages) influence one another during learning? Can the shared syntax, semantics, or programming paradigms across languages be used to improve the generalization capabilities of LLMs? A better understanding of language relationships can inform the construction of more efficient training datasets, improve cross-lingual generalization of LLMs, and guide the creation of programming languages that are more naturally aligned with the strengths and limitations of large models. Furthermore, this knowledge can steer the evolution of programming languages to better meet the needs of both human developers and AI-driven systems.


To this end, we design an embedding-based framework to uncover latent families of 19 widely used programming languages. 
Our core hypothesis is that languages sharing similar features will be positioned closely in the embedding space learned by LLMs, a proposition empirically grounded in NLP \cite{WuYYLK25} and further supported by code LLM studies \cite{Kargaran0YS25, RoziereLCL20}. 
Our framework defines 21 primary linguistic features grounded in fundamental programming elements (\emph{i.e.,} variable definition, branching, loops, and method declaration), and employs LLMs to generate representative code samples for each feature across all languages. Next, the generated code samples are embedded into high-dimensional vectors using a multilingual code LLM. It then clusters feature-aligned vectors for each language and computes pairwise similarities between language embeddings. Finally, it hierarchically clusters the similarity matrix to reveal the family structures among the studied languages.

Our analysis reveals a clear hierarchy of similarities among programming languages, which we refer to as families. Languages sharing grammatical features, such as Java and C++ or Fortran and Pascal, are closer relatives in the embedding space, while languages with heterogeneous linguistic features, such as Python and Java, appear more distant. 
Notably, Go emerges as the centroid language in the embedding space, exhibiting the highest overall similarity to other languages, while Visual Basic aligns most closely with English, consistent with its design philosophy. 

To demonstrate the practical applicability of our findings, we propose three strategies for optimizing multilingual LLM training based on the uncovered linguistic relationships: 
1) \textit{Transfer learning across relatives}, enhancing performance on low-resource languages by fine-tuning with data from closely related high-resource languages;
2) \textit{Curriculum learning for multilingual fine-tuning}, where languages are introduced in an optimal order based on linguistic proximity; and
3) \textit{Identifying universal language} for intermediary code translation tasks, leveraging the centroid properties of languages like Go.
We conduct extensive experiments across four code intelligence tasks. 
The results consistently demonstrate that our embedding-based discovery of programming language families can serve as a foundation for more effective multilingual LLM training.
For example, family-aware transfer learning from Java yields a 19.83\% BLEU-4 score improvement for Swift code summarization. 
Our embedding-distance-guided \textit{Near-to-Far} curriculum learning significantly boosts multilingual code generation performance. 
Furthermore, it significantly improves the intermediary code translation accuracy by leveraging a pivot language strategically positioned between source and target languages in the embedding space.

In summary, our paper makes the following contributions:

\begin{itemize}    
\item We introduce the first automated methodology for discovering programming language families through linguistic similarity analysis, and demonstrate their application in improving multilingual code LLMs. 

\item We develop novel taxonomy-guided strategies for optimal language selection in multilingual code LLM training, including transfer learning and curriculum learning techniques. 

\item We identify \textit{Go} as a centroid programming language, and show its pivotal role in facilitating multilingual code translation with LLMs.

\end{itemize}

\section{Background and Related Work}

\subsection{Taxonomy of Programming Languages} 
A \textit{taxonomy of programming languages} refers to a systematic classification of programming languages based on their features, paradigms, or historical origins, providing a structured way to understand the intricate nature of language. 
While no consensus exists on a unified programming language taxonomy, the academic and industrial community generally organizes them along four key dimensions \cite{pragmatics, kumar2020timeline}: 1) \textit{programming paradigms} delineate computational models, \emph{e.g.}, procedural languages like C and Fortran using step-wise execution, object-oriented languages like Java and C++ employing encapsulation and polymorphism, functional languages like Haskell and Lisp emphasizing pure functions; 2) \textit{execution methods} categorize implementation approaches, \emph{e.g.}, compiled languages like C and C++ generating native machine code, interpreted languages like Python and JavaScript executing through runtime interpretation, and hybrid approaches like Java and C\# utilizing intermediary bytecode compilation; 3) \textit{type systems} characterize type semantics, \emph{e.g.}, static typing in Java and C++ performing compile-time verification, dynamic typing in Python and Ruby deferring type checking to runtime, strong typing in Java and Python enforcing strict constraints, and weak typing in JavaScript and PHP permitting implicit conversions; and 4) \textit{historical lineages} trace language evolution, \emph{e.g.}, the C-family with shared syntax, Java-family for cross-platform development, and the Python family towards data science and AI. 

The advent of LLMs has catalyzed a paradigm shift in programming language taxonomy. LLMs prioritize patterns in code usage, semantic comprehension, syntactic similarity, and task-specific performance~\cite{MaLZXWHZL24, BlackBox}, facilitating a more adaptable and application-centric classification that aligns with their dynamic and context-aware nature. 
Although extensive research has investigated model-based taxonomic analysis for natural languages \cite{TanCHXQL19, MauryaD22, MaLZ23}, the extension of these methodologies to programming language classification constitutes an emerging research frontier.


Our research presents, to our knowledge, the first systematic approach employing LLMs to automatically derive language similarity measures and genealogical connections, providing a scalable, data-driven methodology that surpasses traditional manual or rule-based techniques. In contrast to existing natural language research that primarily relies on multilingual encoder-decoder models with language tag inputs, our approach leverages LLMs to process feature-aligned code snippets as representations of programming languages.

\subsection{Multilingual Learning for Code Intelligence}

Code intelligence employs AI techniques to comprehend and generate source code, facilitating tasks such as code search, program synthesis, and bug detection \cite{WanBHZZSXJY24, allamanis2018survey}. 
Multilingual learning enhances these capabilities across diverse programming languages through shared linguistic patterns, improving model performance and generalization. Research in this area focuses on two primary directions:

\vspace{1.5mm}
\textit{1) Cross-lingual transfer learning}~\cite{Li21, ahmad2021unified} transfers knowledge from source to target languages, particularly effective for low-resource scenarios.
For example, CDCS~\cite{chai2022cross} applies model-agnostic meta-learning to alleviate representation conflicts between source and target languages for code search. 
SDA-Trans~\cite{liu2023syntax} introduces a syntax and domain-aware program translation model that strengthens cross-lingual transfer ability by utilizing syntax structures and domain knowledge. 
IRCoder~\cite{paul2024ircoder} investigates the integration of compiler intermediate representations (IRs) to improve multilingual capabilities in code LLMs and facilitate cross-lingual transfer.

A key challenge in transfer learning is automatically identifying the most beneficial high-resource languages. 
Addressing this issue, Baltaji et al. \cite{baltaji2023learning} conducted empirical research over 41 programming languages and four code-related tasks, analyzing transfer performance across language pairs to establish selection guidelines. 
MIREncoder \cite{dutta2024mirencoder} enhances this process through multi-modal intermediate representations that capture richer syntactic and semantic features.

Our approach differs significantly from theirs. By constructing programming language families with semantic similarity, we establish systematic mechanisms for identifying optimal source languages corresponding to specific targets in transfer learning contexts.

\vspace{1.5mm}
\textit{2) Multilingual training of code LLMs} encompasses model pre-training on heterogeneous multi-language programming corpora to develop cross-lingual code comprehension and synthesis capabilities. 
State-of-the-art code LLMs, such as CodeLlama~\cite{roziere2023code}, DeepSeekCoder~\cite{deepseekcoder2024}, and StarCoder~\cite{li2023starcoder}, usually employ unified architectures for processing multiple languages, yet fundamentally depend on simple corpus aggregation rather than sophisticated cross-lingual learning methodologies.


Recent advancements have introduced more sophisticated multilingual learning frameworks. Baltaji et al.~\cite{baltaji2023learning} proposed a predictive framework for strategic knowledge transfer between high- and low-resource languages; Pian et al.~\cite{pian2023metatptrans} introduced meta-learning to dynamically generate language-specific parameters while preserving shared linguistic features; Paul et al.~\cite{paul2024ircoder} leveraged compiler IRs to mitigate structural and semantic disparities across programming languages.

Unlike existing approaches, our study pioneers a taxonomy-guided framework that optimizes code LLM training through strategic programming language selection and curriculum sequencing.


\section{Uncovering Programming Language Families (RQ1)}

We first explore the linguistic relationships among programming languages and uncover their genealogy characteristics. 

\subsection{Studied Languages}
\label{sec:PLs}

We curate a list of 19 programming languages, including C++, Java, JavaScript, Kotlin, Python, Rust, Haskell, C, Go, Swift, AppleScript, Fortran, Dart, Ruby, Raku, PHP, Visual Basic, Pascal, and Scala. 
These languages are chosen to represent a broad spectrum of programming paradigms such as procedural, object-oriented, functional, and static/dynamic-typing programming. Additionally, the selection balances both high- and low-resource languages to ensure comprehensive coverage. For comparative analysis, English is included as a reference natural language to provide a baseline for linguistic and structural comparisons.

\subsection{Study Design}

Unlike prior studies that rely on human-curated linguistic rules \cite{pragmatics}, we propose an embedding-based framework to uncover latent families of programming languages. Our approach employs LLMs to derive language embeddings and analyzes their genealogical characteristics. Its core rationale is that, through pre-training on large-scale multilingual corpora, LLMs have implicitly acquired knowledge of diverse programming languages. This knowledge enables the encoding of monolingual embeddings within a unified representation space, thereby supporting the study of programming language genealogy.


Figure \ref{fig:RQ1approach} shows the overall design of our study.
We first define key linguistic features for programming languages and craft code samples for each feature by utilizing LLMs. The crafted code samples are then embedded into language vectors using LLMs. Finally, we compute pairwise similarities between the embeddings and cluster programming languages into a hierarchy of families. 
In particular, the pipeline consists of five steps:

\begin{figure*}[t]
       \centering
       \includegraphics[width=0.98\textwidth, trim=0 -5 0 0]{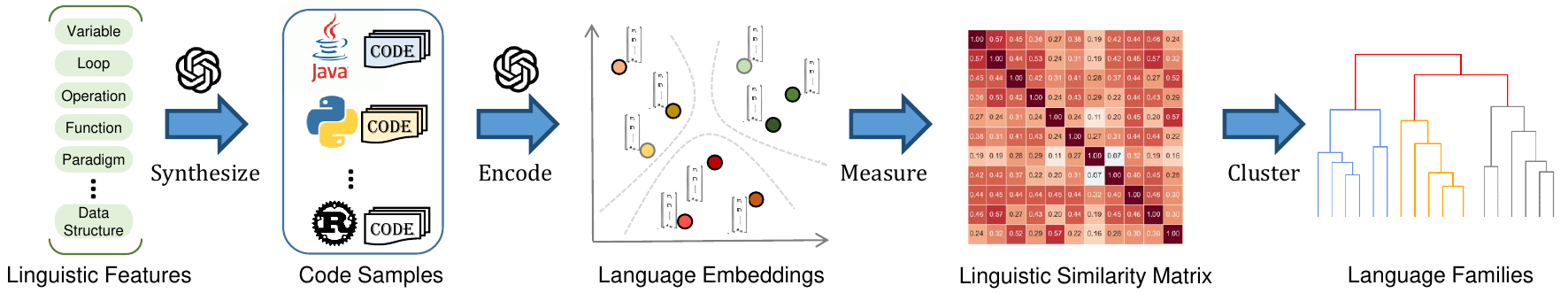}
       \caption{The Workflow for Uncovering Programming Language Families.}
       \label{fig:RQ1approach}
\end{figure*}

\vspace{1.5mm}
1) \textbf{Defining Linguistic Features}. 
To characterize programming languages, we design 21 linguistic features grounded in fundamental programming elements, \emph{e.g.}, variable definition, typing, branching, loops, and method declaration. 
These features are initially collected from language references provided in the official documentation of each language\footnote{https://docs.python.org/3/reference/index.html}\footnote{https://docs.oracle.com/javase/specs/jls/se24/html/index.html}\footnote{https://en.cppreference.com/w/c/language.html}\footnote{https://docs.ruby-lang.org/en/master/}\footnote{https://devdocs.io/javascript/}.
To ensure robustness and mitigate individual bias, three programming language experts from our institution first independently synthesize these elements into a preliminary feature set. This initial set is then systematically reviewed by two authors. Each proposed feature is rigorously evaluated against two core criteria: its technical justification as a fundamental and distinguishing aspect of a programming paradigm or syntax, and its cross-language expressibility, meaning the concept must be implementable in all languages under study. Each feature is supported by a technical rationale and refined iteratively through structured discussion among all participants. The process continues until consensus is reached and all methodological concerns are resolved. 

\begin{table}[t]
  \centering
  \caption{The 21 Linguistic Features of Programming Languages}
  \small
  \begin{tabularx}{\linewidth}{c >{\raggedright\arraybackslash}p{0.28\linewidth} Y}
    \toprule
    \textbf{Feature} & \textbf{Name} & \textbf{Description} \\
    \midrule
    $F_1$ & Variable Definition & How does a language define variables of various types, particularly in distinguishing static and dynamic typing? \\
    \rowcolor{gray!30}
    $F_2$ & Conditional Branching & How does a language realize conditions and branches in program control flow? \\
    $F_3$$\sim$$F_4$ & Loop: For and While & How does a language implement loop constructs in program control flow? \\
    \rowcolor{gray!30}
    $F_5$ & System I/O & How does a language handle standard input and output operations, such as reading user input and printing text to the screen? \\
    $F_6$$\sim$$F_8$ & Operations: Arithmetic, Logical, and Comparison & Basic features governing operations, covering syntax, hierarchy, and conditional evaluation.\\
    \rowcolor{gray!30}
    $F_{9}$ & Library Integration & How does a language import and utilize standard and third-party libraries? \\
    $F_{10}$ & Parameter Passing & What are the mechanisms for passing arguments in function calls, including distinctions between pass-by-value, pass-by-reference, and other strategies? \\
    \rowcolor{gray!30}
    $F_{11}$ & Function Returns & How does a language define and manage return values from functions, including support for multiple return values or return type declarations? \\
    $F_{12}$ & Exception Handling & How does the language manage runtime errors, including syntax and semantics of exception-throwing and catching constructs? \\
    \rowcolor{gray!30}
    $F_{13}$$\sim$$F_{16}$ & Data Structures: Array, List, Set, and Map & What are the built-in data abstraction mechanisms provided by the language, such as arrays, lists, sets, and maps, and how are they typically used? \\
    $F_{17}$$\sim$$F_{19}$ & OOP: Class Definition, Object Creation, Inheritance & How does the language support object-oriented constructs such as class definitions, object instantiation, encapsulation, inheritance, and polymorphism? \\
    \rowcolor{gray!30}
    $F_{20}$$\sim$$F_{21}$ & Functional Programming: Map and Filter & A declarative paradigm emphasizing pure functions and immutable data, with Map and Filter exemplifying key features. \\
    \bottomrule
  \end{tabularx}
  \label{tab:feature}
\end{table}

Table~\ref{tab:feature} summarizes the final validated feature set.
Among them, the feature \textit{variable declaration} reflects Python's dynamic typing paradigm, wherein type inference occurs upon assignment without requiring explicit type annotations. The \textit{conditional branching} feature captures Haskell's \textit{guard} construct, which streamlines conditional logic by substituting nested \texttt{if-then-else} structures with a pipe symbol (\texttt{|}) preceding Boolean expressions. Features $F_{10}$ and $F_{11}$ encode C's function invocation properties, such as mandatory return type specifications in function signatures, named functions with parenthesized parameter lists, and void functions that omit return values. 
To represent object-oriented paradigms, we incorporate core OOP features such as class definitions, object instantiation, and inheritance mechanisms \cite{meyer1997object}. For functional programming, we focus on features of higher-order functions such as implementations of \texttt{map} and \texttt{filter}. They represent immutability and function composition, two core principles in functional programming \cite{harrison1997introduction}.

\begin{table*}[t]
\centering
\caption{Examples of Curated Code Samples for Linguistic Features} 
\label{tab:GeneratingCode}
\begin{tabular}{c|c|c|c|c}
\toprule
\scriptsize \textbf{Feature} 
& {\scriptsize \textbf{Python}}
& {\scriptsize \textbf{PHP}}
& {\scriptsize \textbf{Ruby}}
& {\scriptsize \textbf{Dart}} \\
\hline
\scriptsize \makecell{Variable\\Definition}
& \raisebox{-.5\height}{\includegraphics[width=2.5cm]{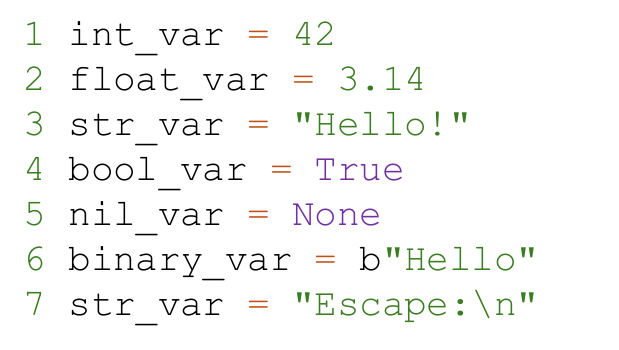}}
& \raisebox{-.5\height}{\includegraphics[width=2.5cm]{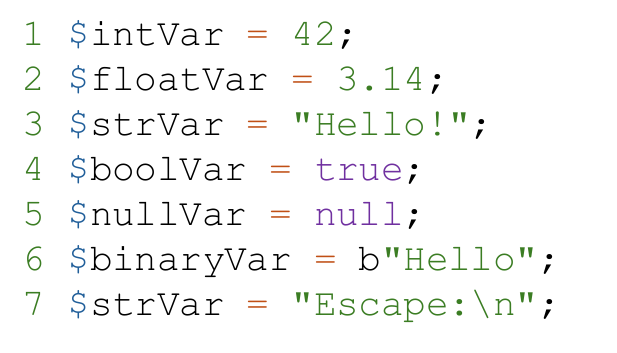}}
& \raisebox{-.5\height}{\includegraphics[width=2.5cm]{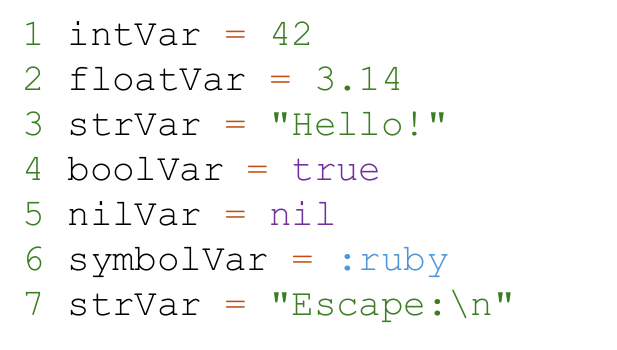}}
& \raisebox{-.5\height}{\includegraphics[width=2.5cm]{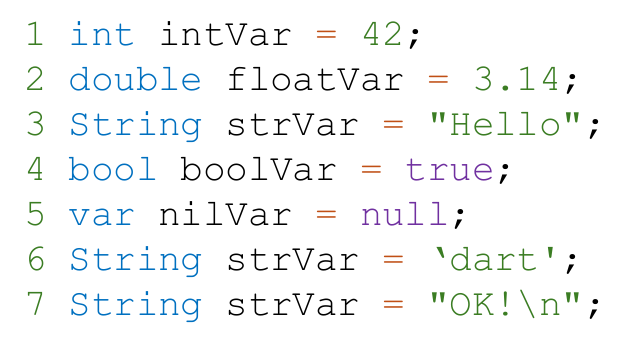}}
\\
\hline
\scriptsize \makecell{Conditional\\Branching}
& \raisebox{-.5\height}{\includegraphics[width=2.5cm]{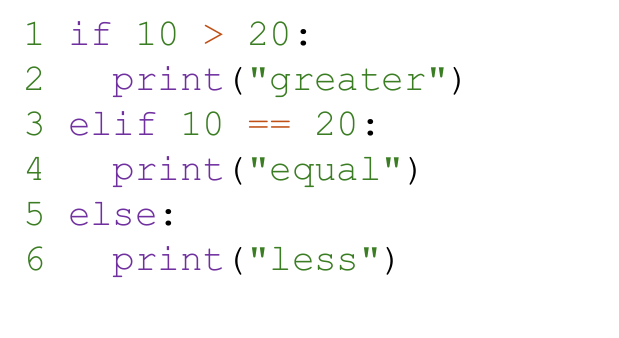}}
& \raisebox{-.5\height}{\includegraphics[width=2.5cm]{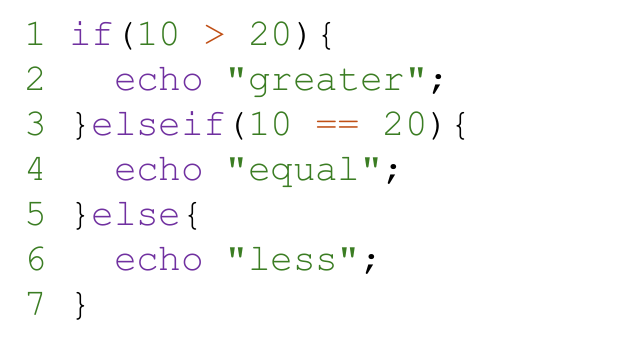}}
& \raisebox{-.5\height}{\includegraphics[width=2.5cm]{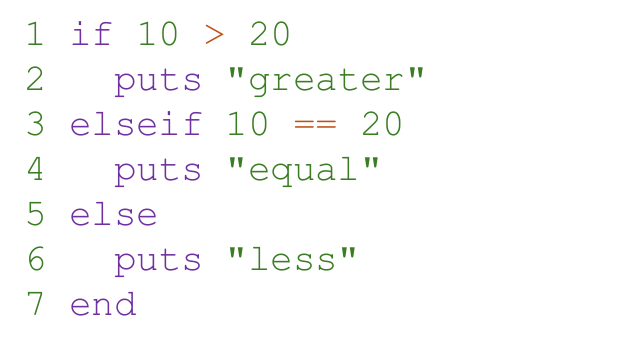}}
& \raisebox{-.5\height}{\includegraphics[width=2.5cm]{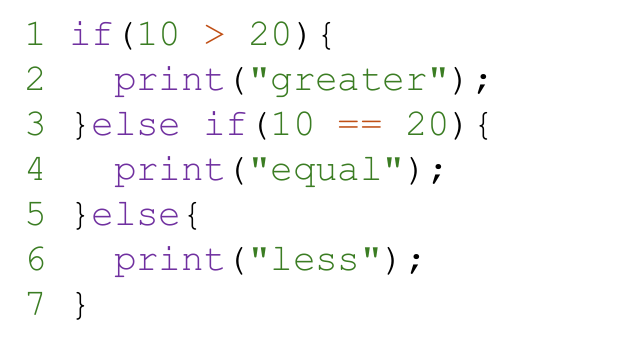}}
\\
\hline
\scriptsize \makecell{For\\Loop}
& \raisebox{-.5\height}{\includegraphics[width=2.5cm]{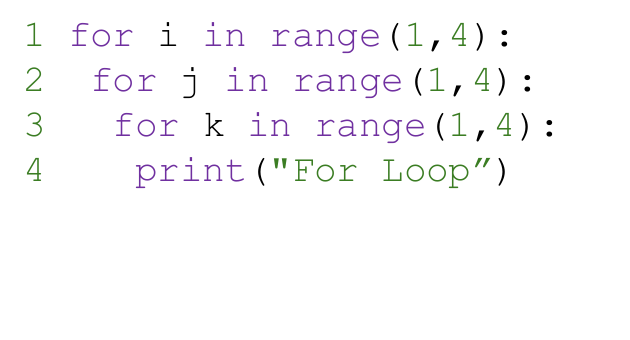}}
& \raisebox{-.5\height}{\includegraphics[width=2.5cm]{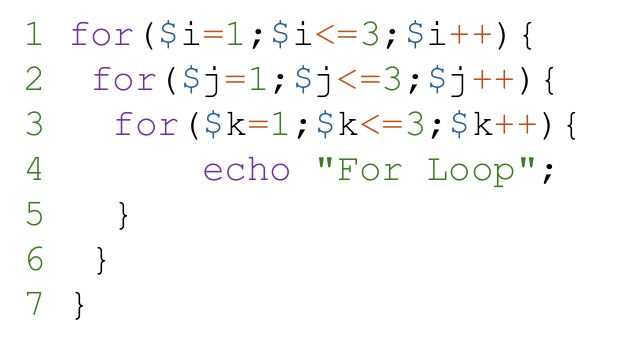}}
& \raisebox{-.5\height}{\includegraphics[width=2.5cm]{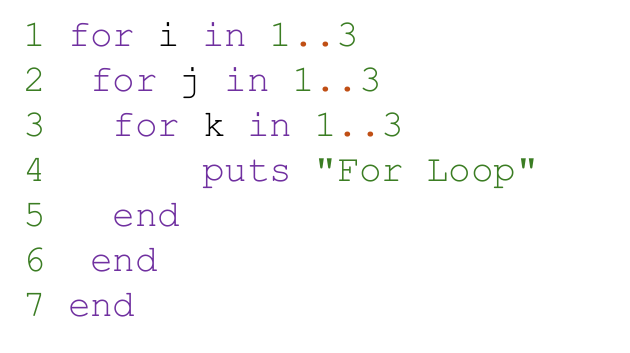}}
& \raisebox{-.5\height}{\includegraphics[width=2.5cm]{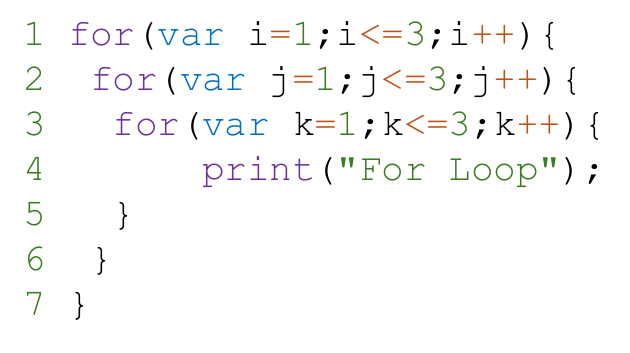}}
\\
\hline
\scriptsize \makecell{Arithmetic\\Operations}
& \raisebox{-.5\height}{\includegraphics[width=2.5cm]{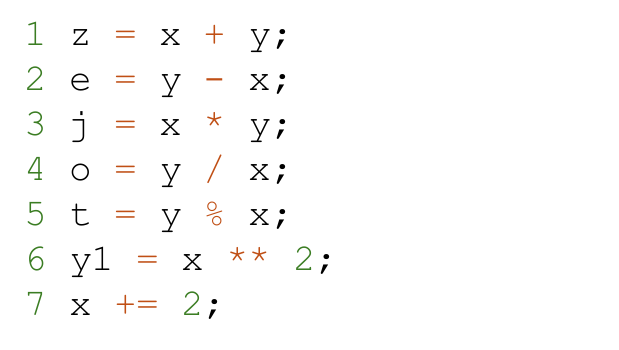}}
& \raisebox{-.5\height}{\includegraphics[width=2.5cm]{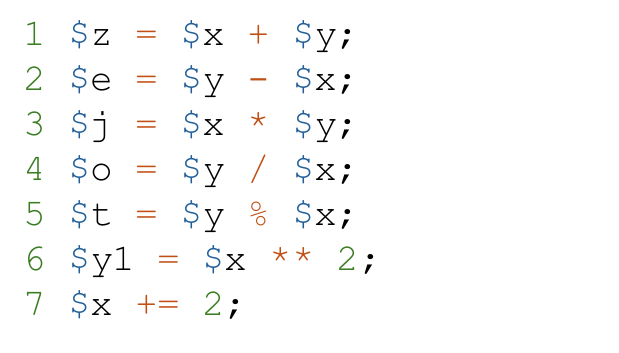}}
& \raisebox{-.5\height}{\includegraphics[width=2.5cm]{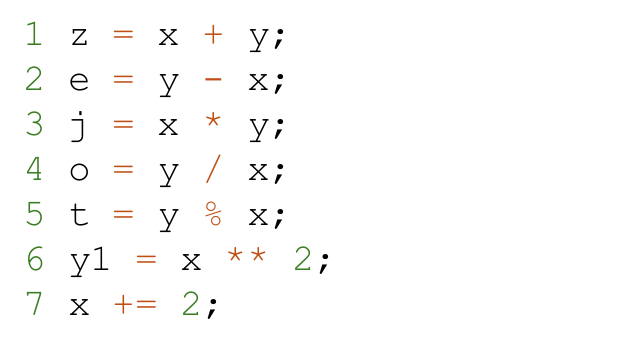}}
& \raisebox{-.5\height}{\includegraphics[width=2.5cm]{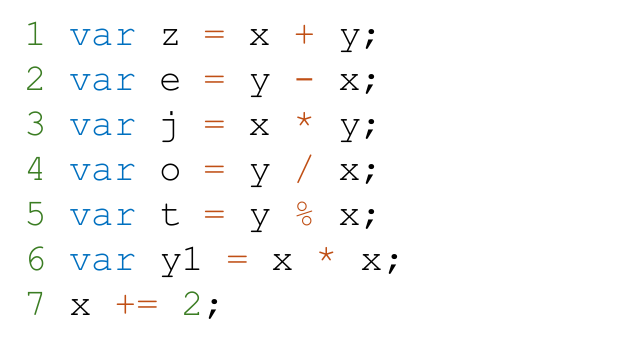}}
\\
\hline
\scriptsize \makecell{Object\\Oriented}
& \raisebox{-.5\height}{\includegraphics[width=2.5cm]{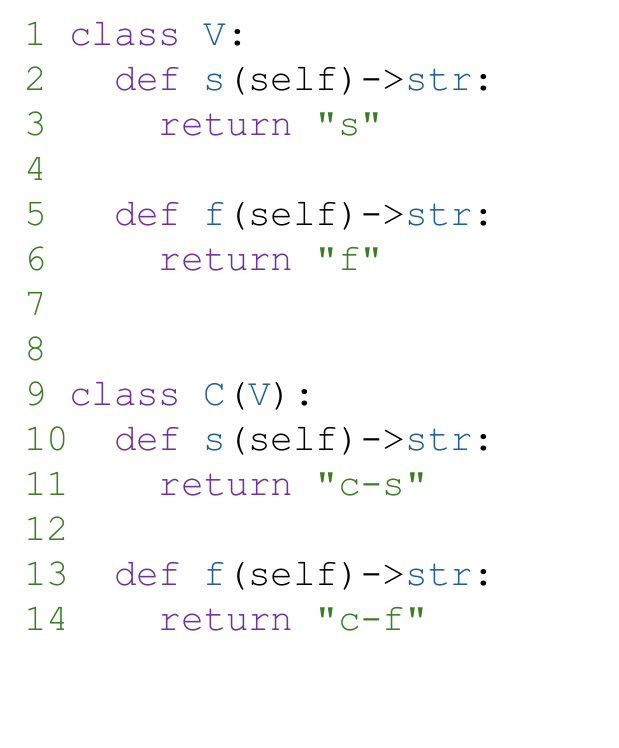}}
& \raisebox{-.5\height}{\includegraphics[width=2.5cm]{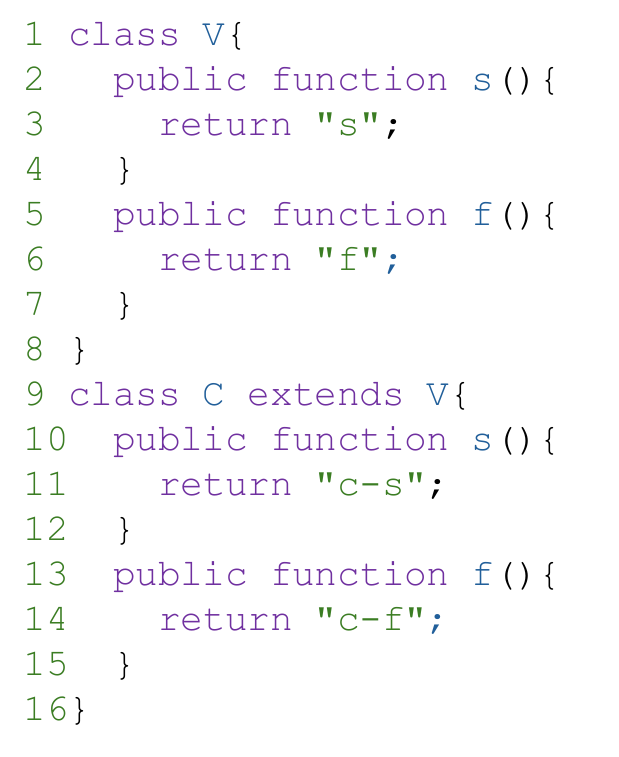}}
& \raisebox{-.5\height}{\includegraphics[width=2.5cm]{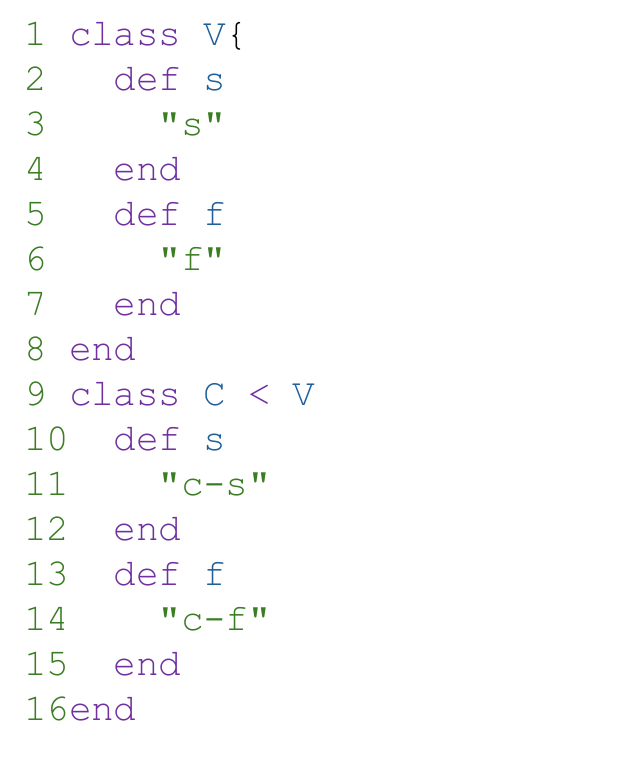}}
& \raisebox{-.5\height}{\includegraphics[width=2.5cm]{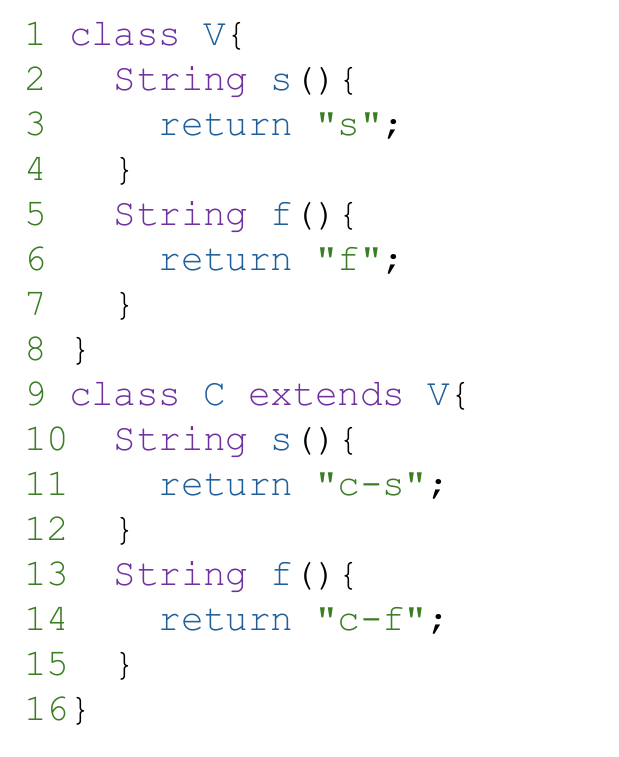}}
\\
\hline
\scriptsize \makecell{Functional\\Programming}
& \raisebox{-.5\height}{\includegraphics[width=2.5cm]{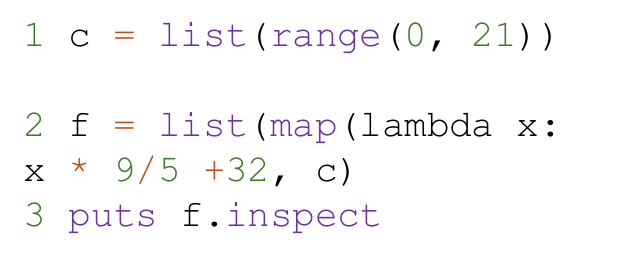}}
& \raisebox{-.5\height}{\includegraphics[width=2.5cm]{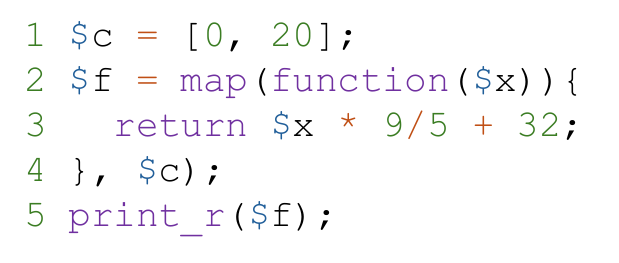}}
& \raisebox{-.5\height}{\includegraphics[width=2.5cm]{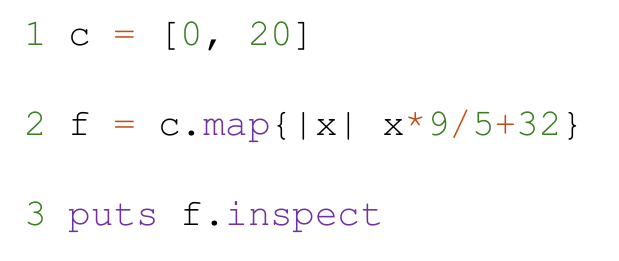}}
& \raisebox{-.5\height}{\includegraphics[width=2.5cm]{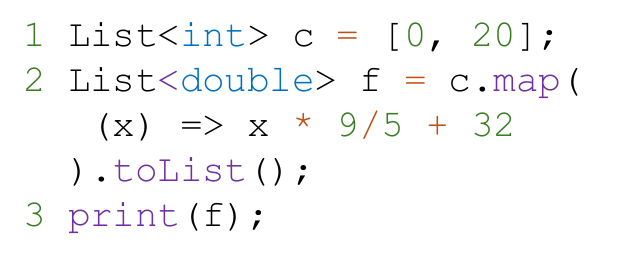}}
\\
\bottomrule
\end{tabular}
\end{table*}

\vspace{1.5mm}
2) \textbf{Curating Code Samples}.
For each linguistic feature, we curate code samples that instantiate each primary linguistic feature. These code samples can be subsequently embedded using LLMs to derive vectorized feature representations. We rely on LLM-generated code rather than samples sourced from real-world projects due to the need for a large-scale corpus of parallel-semantic code snippets across 19 programming languages. Sourcing such a balanced and diverse dataset from existing codebases would be highly challenging and likely infeasible. The use of LLMs enables the generation of linguistically consistent and controlled samples across all languages, which is critical for ensuring cross-lingual comparability. 

More specifically, we instruct GPT-4o~\cite{GPT-4o} to generate a parallel code corpus comprising 1,900 samples per feature across 19 programming languages using a structured prompt. The resulting corpus contains 39,900 code snippets in total, with each language represented by 2,100 uniformly distributed samples.

\newcounter{prompt}
\begin{tcolorbox}[colframe=darkgray2, colback=white, coltitle=white, colbacktitle=darkgray2, 
breakable,
title=Prompt Template for Generating Feature-aligned Code Samples, 
boxrule=0.8pt,
fonttitle=\mdseries\small, fontupper=\ttfamily\footnotesize, , rounded corners]
\refstepcounter{prompt}
Produce code exemplars for \textcolor{blue}{\$FEATURE\_NAME} in C++, Java, JavaScript, Kotlin, Python, Rust, Haskell, C, Go, Swift, AppleScript, Fortran, Dart, Ruby, Raku, PHP, Visual Basic, Pascal, and Scala.\\
\# \textcolor{blue}{\$FEATURE\_NAME}: \textcolor{violet}{\$FEATURE\_DESCRIPTION}\\
\# You must strictly adhere to the following rules:\\
1) Generate 100 code snippets for each language;\\
2) These 100 code snippets must not only conform to the feature specification but should also be maximally diversified;\\
3) Ensure semantic consistency across code snippets in different languages, meaning they should implement the same functionality.
\end{tcolorbox}

\new{To ensure quality, we perform manual spot checks by three programming language experts. They examine 10\% of the samples, evaluating each based on four criteria: adherence to feature specifications, maximal diversity, syntactic validity, and functional consistency. If non-compliant instances are identified, the experts expand the inspection scope and instruct the LLM to regenerate or correct the outputs until all criteria are satisfied.}
Table~\ref{tab:GeneratingCode} showcases several samples of generated code for six representative features (\textit{variable declaration}, \textit{branching}, \textit{loop}, \textit{arithmetic operations}, \textit{object-oriented} and \textit{functional programming}), across Python, PHP, Ruby, and Dart. 


\begin{figure*}[t]
    \centering
    \includegraphics[width=\textwidth, trim=0 10 0 0]{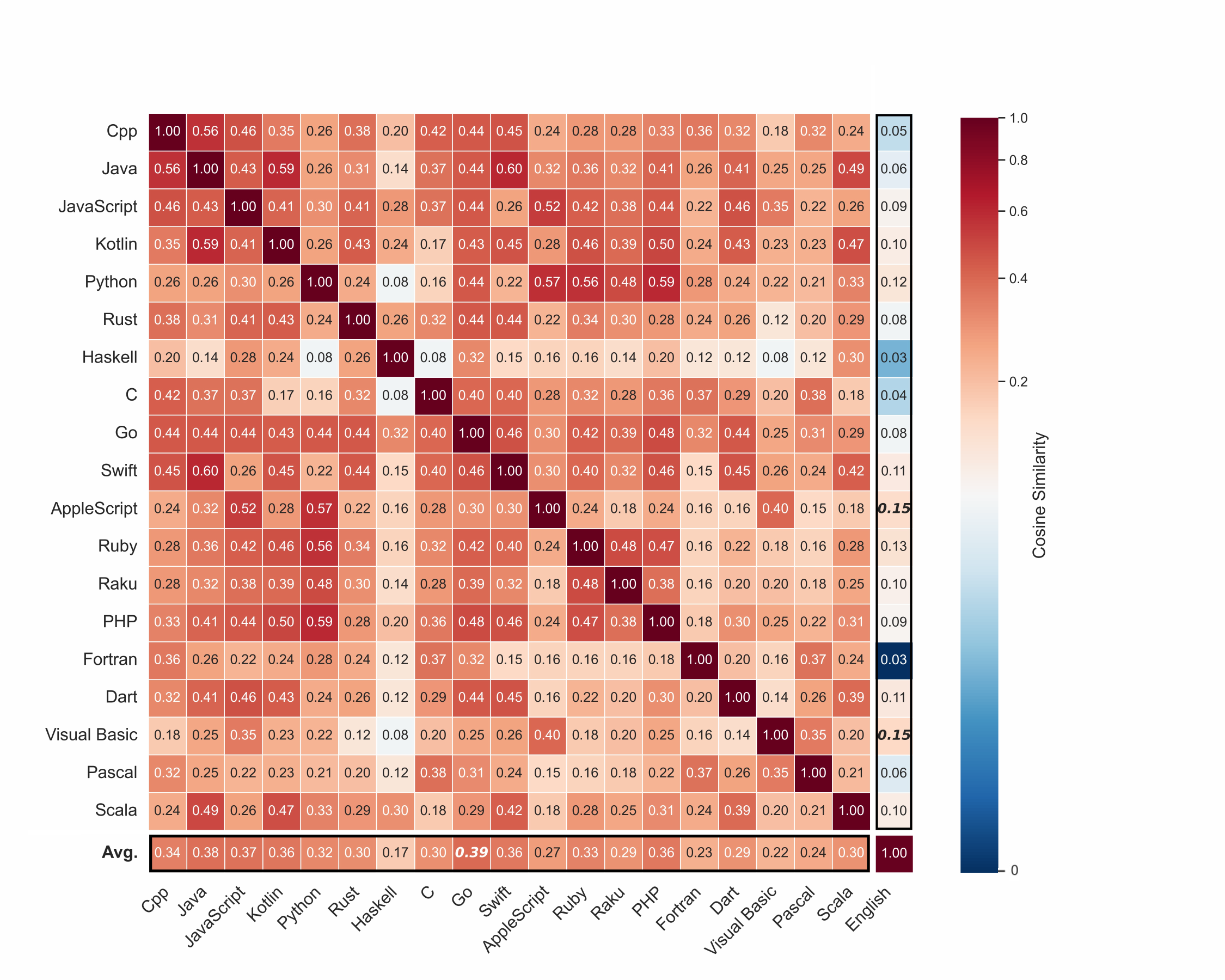}
    \caption{A Heatmap of Pairwise Similarities across Programming Languages. A deeper shade of red indicates a higher similarity. The bottom row represents the mean similarity to other languages, excluding English.}
    \label{fig:matrix}
\end{figure*}

\vspace{1.5mm}
3) \textbf{Language Embedding}. 
We encode the curated code samples into high-dimensional vectors to capture their linguistic characteristics. To this end, we employ LLM embedding model such as OpenAI's text-embedding\footnote{https://openai.com/index/new-and-improved-embedding-model/} to compute feature-specific embeddings for each programming language. 
Each feature vector corresponds to the centroid of its associated code vectors, and the final language embedding is derived by aggregating all feature vectors of the corresponding language. This parallel corpus approach effectively preserves language-specific distinctions within the embedding space.

\vspace{1.5mm}
4) \textbf{Similarity Analysis}. 
Within the shared multilingual embedding space, we quantitatively assess inter-language relationships through pairwise similarity analysis. The normalized cosine similarity between language embeddings $\mathcal{L}_a$ and $\mathcal{L}_b$ is computed as:
\begin{align}
\text{Similarity}(\mathcal{L}_a, \mathcal{L}_b) = \frac{1}{2} + \frac{1}{2}\cos(\mathbf{v}_a, \mathbf{v}_b)
\end{align}
where $\mathbf{v}_a$ and $\mathbf{v}_b$ denote the embedding vectors of $\mathcal{L}_a$ and $\mathcal{L}_b$, respectively, and $\cos(\mathbf{v}_a, \mathbf{v}_b)$ denotes the conventional cosine similarity metric (range: [-1,1]). This transformation ensures the similarity measure is bounded within the [0,1] interval, maintaining both interpretability and alignment with established research conventions.


\vspace{1.5mm}
5) \textbf{Clustering Languages}. 
Finally, we perform hierarchical clustering~\cite{TanCHXQL19} using Ward's minimum variance algorithm~\cite{ward1963hierarchical} to analyze programming language genealogy. 
The clustering process operates on the pairwise language similarity matrix to construct a dendrogram of linguistic relationships.  
The optimal cluster count \( K \) is automatically determined through the elbow criterion~\cite{elbow}, revealing inherent groupings that reflect linguistic affinities among programming languages.



\subsection{Results and Analysis} 
\label{sec:matrix}

\textbf{Cross-Language Similarity Analysis}.
Figure~\ref{fig:matrix} presents the pairwise similarity matrix for 19 programming languages. 
We observe that mainstream languages, including C, C++, Java, JavaScript, and Go, exhibit close proximity in the embedding space, with similarity scores spanning 0.37$\sim$0.56 ($\mu$=0.43, $\sigma$=±0.05). This convergence reflects their shared syntactic structures, control flow paradigms, and library implementations in contemporary software development.

Conversely, Haskell and Fortran demonstrate significantly lower similarity compared to other languages, with mean coefficients of merely 0.17 and 0.23 respectively, revealing their unique representational properties. This observed dissimilarity originates from their unique computational paradigms and linguistic implementations.
The purely functional paradigm of Haskell, characterized by lazy evaluation, pattern matching, and higher-order functions, establishes fundamental differences from mainstream imperative languages. Correspondingly, Fortran, as a pioneering scientific computing language, exhibits a fundamentally distinct syntax and data structure architecture specifically optimized for numerical computation, resulting in substantial architectural differences from contemporary general-purpose languages.

\begin{tcolorbox}[enhanced, width=\linewidth, boxrule=0.8pt, 
 left=2pt, right=2pt, top=2pt, bottom=2pt, drop fuzzy shadow=black,]
\textbf{Finding 1:} 
Mainstream languages, such as \textit{C} and \textit{Java}, exhibit close proximity in the embedding space while paradigm-specific languages occupy isolated positions.
\end{tcolorbox}
\vspace{2pt}

\textbf{Average Similarity}. 
The similarity matrix's final row quantifies each language's mean similarity to others. Go achieves the highest cross-lingual similarity (0.39), occupying the geometric center of the embedding space. 
This centrality emerges from three intrinsic linguistic properties of Go: lexical universality across languages, reduced semantic complexity compared to polysemous constructs, and stable cross-lingual representational alignment.
This finding aligns with existing research~\cite{TaoYGS24} that identified Go as exhibiting maximal LLM interpretability across comparative language analyses.

Java occupies the second most central position with a semantic similarity of 0.38 to other languages, indicating strong semantic alignment with mainstream programming paradigms.
In contrast, Haskell demonstrates maximal linguistic divergence (0.17) from other languages, due to its purely functional paradigm's unique computational model and representational characteristics. 

\begin{tcolorbox}[enhanced, width=\linewidth, boxrule=0.8pt, 
 left=2pt, right=2pt, top=2pt, bottom=2pt, drop fuzzy shadow=black,]
\textbf{Finding 2:} 
\textit{Go} has maximal cross-lingual semantic similarity among programming languages, occupying the geometric centroid of the multilingual embedding space.
\end{tcolorbox}
\vspace{2pt}

\textbf{Relationships with English}. 
The rightmost column of the similarity matrix measures the pairwise similarity coefficients between each programming language and English.
As anticipated, English occupies the outermost periphery of the representation space, showing minimal linguistic similarity ($\mu$=0.088) to programming languages due to fundamental structural disparities between natural and formal languages.  
Among the studied programming languages, Haskell and Fortran exhibit maximal divergence from English ($\mu$=0.03), reflecting their specialized computational paradigms. Conversely, Visual Basic and AppleScript display higher syntactic alignment with English, resulting from deliberate design decisions prioritizing natural language-like readability.

\begin{tcolorbox}[enhanced, width=\linewidth, boxrule=0.8pt, 
 left=2pt, right=2pt, top=2pt, bottom=2pt, drop fuzzy shadow=black,]
\textbf{Finding 3:} 
\textit{Visual Basic} shares the most similar syntax with English, while \textit{Haskell} and \textit{Fortran} exhibit the most distinct structural differences from natural language paradigms.
\end{tcolorbox}
\vspace{2pt}

\begin{figure}[t]
       \centering
       \hspace*{-1em}
       \includegraphics[width=0.78\textwidth, trim=0 0 0 0]{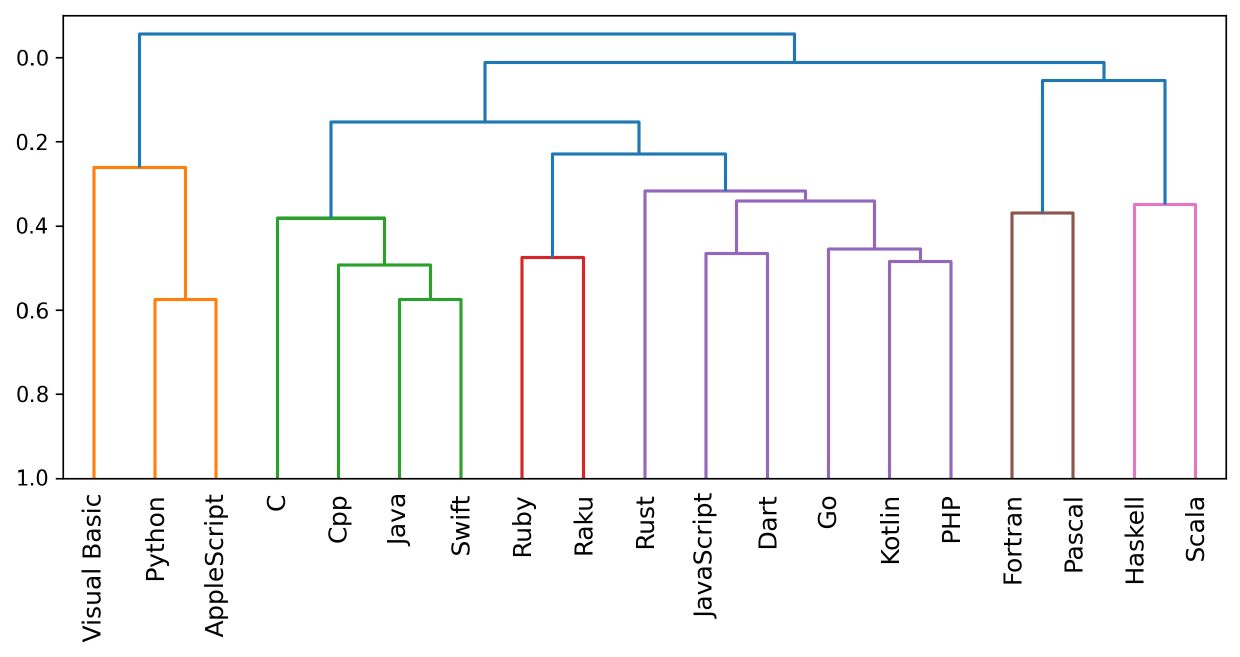}
       \caption{The Hierarchical Clustering of Programming Languages. The Y-axis represents the similarity between two languages or clusters. Languages in the same color are divided into the same cluster, where blue color agglomerates different clusters together.
       }
       \label{fig:cluster}
\end{figure}

\textbf{Linguistic Clusters (Families)}.
We perform hierarchical clustering analysis on the pairwise language similarity matrix, and the results are visualized in Figure~\ref{fig:cluster}. 
Overall, we have identified six language families, characterized by their semantic traits, typing systems, programming paradigms, syntactic structures, and typical use cases.

\textit{Cluster 1: The C Family}, comprising C, C++, Java, and Swift. This group shares syntactic similarities, static typing, and adherence to the imperative programming paradigm. These languages exhibit a common historical lineage, as C++ evolved directly from C, while Java and Swift have extensively incorporated syntactical elements and design philosophies from both C and C++.

\textit{Cluster 2: Modern Multi-Paradigm Languages}, containing Rust, JavaScript, Dart, Go, Kotlin, and PHP. These languages emphasize development efficiency through syntactic conciseness, type flexibility, and sophisticated concurrency models (\emph{e.g.}, Goroutines, async/await). Their paradigm-blending designs combine functional, procedural, and object-oriented approaches.

\textit{Cluster 3: Scripting Languages}, including Visual Basic, Python, and AppleScript, featuring natural language-inspired syntax and dynamic typing. These languages employ interpreter-based execution, optimizing them for scripting and automation tasks through flexible variable usage and rapid prototyping capabilities.


\textit{Cluster 4: Ruby \& Raku}, emphasizing dynamic, expressive syntax and meta-programming capabilities.

\textit{Cluster 5: Fortran \& Pascal}, serving numerical computation and structured programming education respectively.

\textit{Cluster 6: Haskell \& Scala}, distinguished by advanced functional programming features and rigorous type systems.

\new{To assess the structural stability of the derived language families, we compute cluster validation indices. The resulting silhouette coefficient of 0.41 indicates clear separability and meaningful organization among the clusters.}
These clustering results effectively reveal fundamental organizational principles across programming languages, validating our embedding method's ability to capture linguistic relationships.

\begin{tcolorbox}[enhanced, width=\linewidth, boxrule=0.8pt, 
 left=2pt, right=2pt, top=2pt, bottom=2pt, drop fuzzy shadow=black,]
\textbf{Finding 4:}  
Our embedding-driven analysis reveals six linguistically coherent programming language clusters, differentiated by their type systems, computational paradigms, syntactic structures, and domain-specific implementations. 
\end{tcolorbox}
\vspace{2pt}

\textbf{Correlations with Existing Taxonomies}. 
The clusters identified through our analysis show substantial alignment with conventional programming language taxonomies.  
For example, the C-family cluster (C, C++, Java, Swift) maintains cohesion through shared syntactic conventions, static typing regimes, and explicit control flow structures.  
Scripting languages like Visual Basic, Python, and AppleScript form a cohesive cluster owing to their shared characteristics of dynamic typing, interpreter-based execution, and developer productivity focus.
Similarly, the Haskell-Scala cluster reflects their foundational functional programming principles. Despite Scala's multi-paradigm flexibility and JVM integration, both languages consistently exhibit strong static typing, immutable data structures, and declarative programming models that our methodology accurately represents.  

These family structures are further corroborated by prior neuron-based analyses of LLMs. Through techniques including neuron activation tracking and cross-language representation alignment, Kargaran et al. \cite{Kargaran0YS25} found that C-family languages (C, C++, C\#) and Java cluster closely in internal model representations, while Python and PHP align more closely with JavaScript. 
These findings are consistent with our hierarchical clustering results, providing external validation that our embedding-based approach captures linguistically meaningful features which resonate with both human-defined classifications and the intrinsic structures within language models.

Notably, our taxonomy extends beyond replicating existing categorizations by reflecting functional linguistic traits and modern usage patterns directly inferred from LLM embeddings. It reveals novel groupings that better reflect the operational characteristics and pragmatic demands of contemporary development practice. A compelling example is the consistent grouping of Rust, JavaScript, Dart, Go, Kotlin, and PHP, despite variations in their type systems (static vs. dynamic), execution models (compiled vs. interpreted), and supported paradigms. This convergence may be attributed to shared emphases on modern development productivity, sophisticated concurrency support, and dynamic programming capabilities.


\begin{tcolorbox}[enhanced, width=\linewidth, boxrule=0.8pt, 
 left=2pt, right=2pt, top=2pt, bottom=2pt, drop fuzzy shadow=black,]
\textbf{Finding 5:}  
The families uncovered through language embedding analysis demonstrate strong correlations with conventional taxonomies and model neuron findings in representing fundamental semantic characteristics, yet reveal variations in languages' practical usage patterns.
\end{tcolorbox}
\vspace{2pt}

\new{It should be noted that our approach is agnostic to the specific embedding model used. To compute feature-specific embeddings for each programming language, we employed three distinct models: OpenAI's text-embedding-ada-002, BGE, and UniXCoder. Despite differences in their training objectives and architectures, all models produced highly similar clustering structures. These results suggest that the discovered language families are robust.}

\new{Furthermore, our pipeline is computationally efficient. The code synthesis step for 39.9k snippets required 51 hours at a cost of \$83, while embedding and clustering together took less than 10 minutes on a standard workstation. The design also scales linearly with the number of languages and features, enabling straightforward expansion to larger typological spaces with minimal engineering overhead.}

\section{Implications for Code LLM Training and Inference (RQ2)} 

\begin{figure}[t]
       \centering         \includegraphics[width=0.85\textwidth, trim=0 -4 0 0]{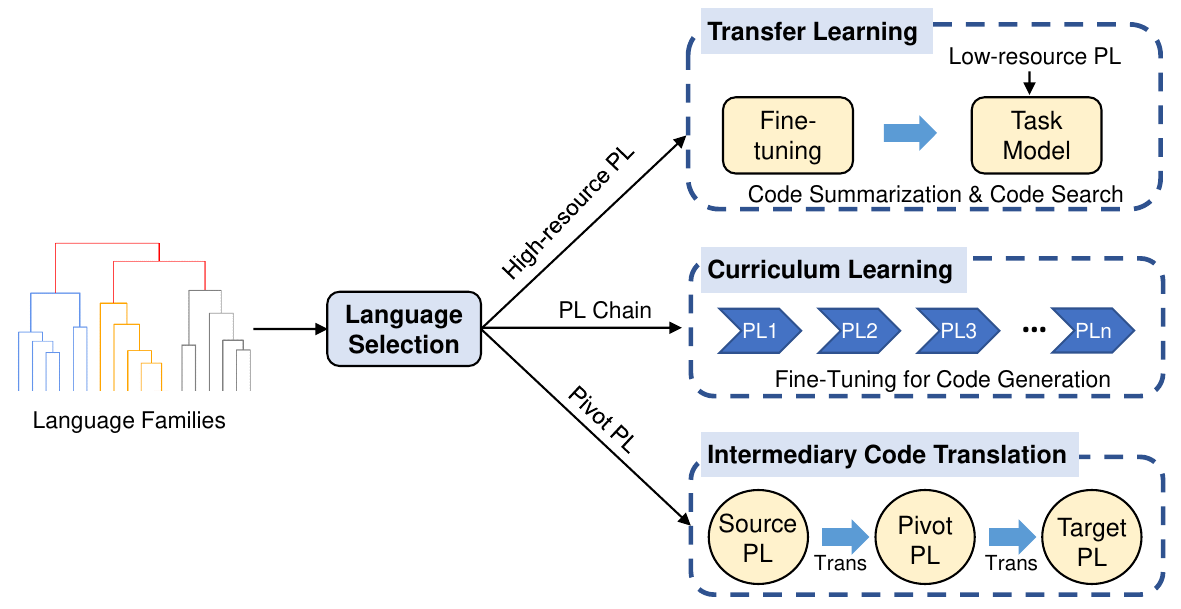}
       \caption{Implications of Programming Language Families for Code LLM Training and Inference.}
       \label{fig:RQ2approach}
\end{figure}

To validate our findings and illustrate the practical applicability of the uncovered language families in code LLM development, we design and conduct a series of targeted experiments. \new{We select three representative scenarios as proof-of-concept demonstrations, each assigned one or two specific tasks}, as shown in Figure~\ref{fig:RQ2approach}. Through four code intelligence tasks (i.e., code summarization, search, generation, and translation), we evaluate whether the proposed language taxonomy can offer guidance for language selection and training strategies.

Specifically, we investigate the following research sub-questions:

\begin{itemize}    
\item\textbf{RQ2.1} \textit{Do language families facilitate transfer learning between high- and low-resource programming languages? (Section \ref{sec:transfer})}

\item\textbf{RQ2.2} \textit{Can language families optimize multilingual LLM fine-tuning through strategic curriculum design? (Section \ref{sec:curriculum})}

\item\textbf{RQ2.3} \textit{Does language family analysis enable the identification of optimal pivot programming languages for intermediary code translation? (Section \ref{sec:translation})}
\end{itemize}

\new{This design allows us to illustrate the utility of our taxonomy across diverse tasks and learning paradigms, confirming that the findings are generalizable and applicable to broader contexts. }

\subsection{Transfer Learning for Low-resource Languages}
\label{sec:transfer}

Prior studies have shown that fine-tuning LLMs on a high-resource language can enhance the performance of low-resource languages \cite{guo2022unixcoder}. A practical challenge of this methodology is how to determine the optimal high-resource languages~\cite{abs-2310-16937}.
Our study provides a new insight into this problem, that is, we can fine-tune code LLMs on a high-resource language that is most similar to the low-resource language in the language family.
To verify the effectiveness of this strategy, we fine-tune code LLMs on Java and Python, two high-resource languages, respectively, and transfer the knowledge to four low-resource languages (Kotlin, Haskell, Swift, and AppleScript), respectively. Extensive experiments are performed across all source-target language pairs. 

\vspace{1.5mm}
\textbf{Experimental Setup.} We evaluate the performance through two code-intelligent tasks: code summarization and code search. The performance is measured by BLEU-4~\cite{bleu} (for summarization) and Top-10 accuracy~\cite{lu2021codexglue} (for search), two respective metrics for these tasks. 
We choose StarCoderBase-15.5B~\cite{starcoder} as our base model due to its publicly available training dataset statistics, which help distinguish between high- and low-resource programming languages.
To mitigate data leakage, we construct a new dataset by scraping GitHub repositories across six programming languages, yielding over 9,000 code-description pairs. We apply a temporal filtering strategy, ensuring test samples are sourced exclusively from repositories updated after StarCoderBase-15.5B’s training cutoff date.
The dataset is partitioned into two subsets: a fine-tuning subset containing 1,500 parallel code-description pairs for Java and Python, and a testing subset comprising 700 samples each for Kotlin, Haskell, Swift, and AppleScript. 

\begin{figure}[t]
  \centering
  \begin{subfigure}{.48\linewidth}
    \centering
    \includegraphics[width=\linewidth]{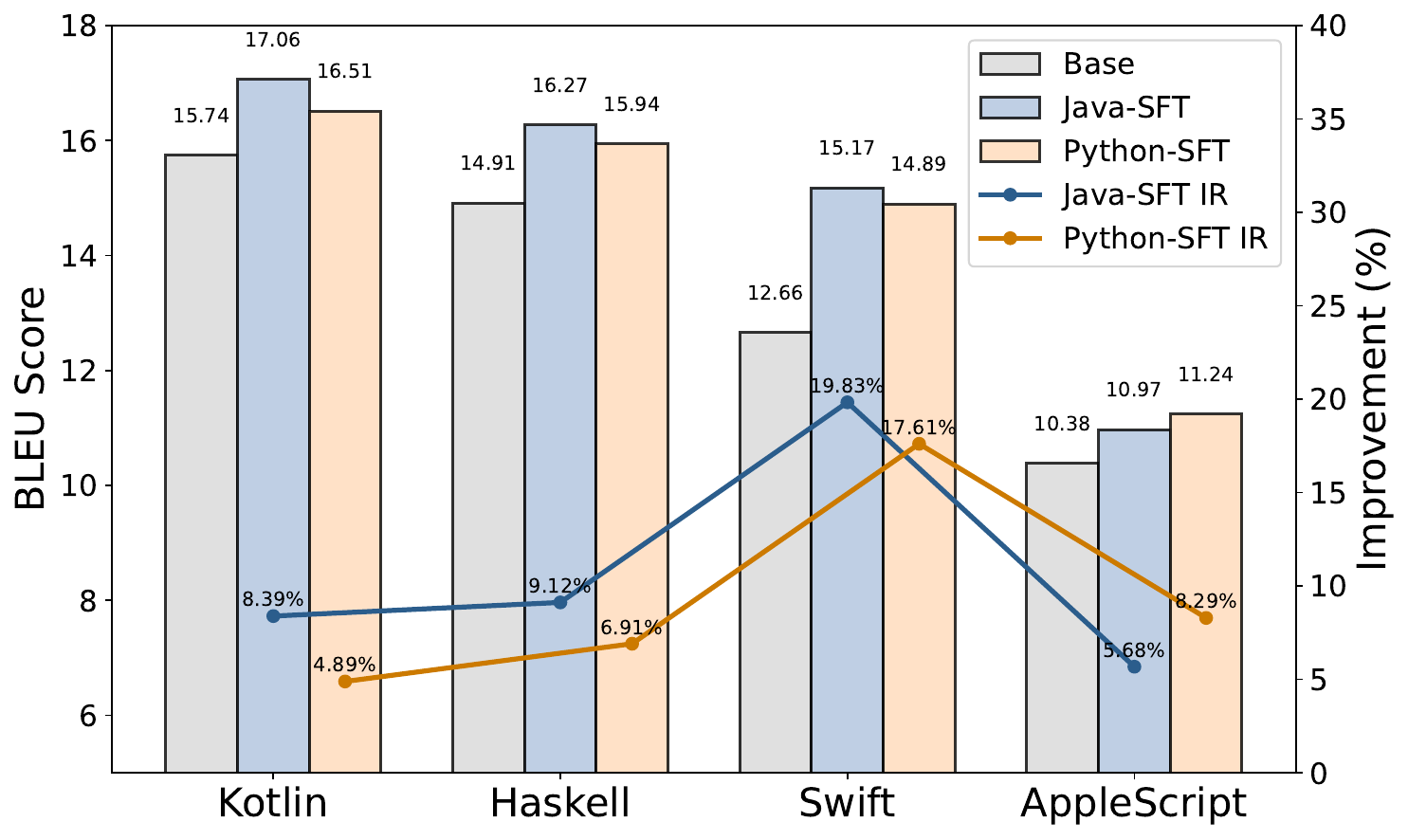}
    \caption{Code Summarization}
  \end{subfigure}\hfill
  \begin{subfigure}{.48\linewidth}
    \centering
    \includegraphics[width=\linewidth]{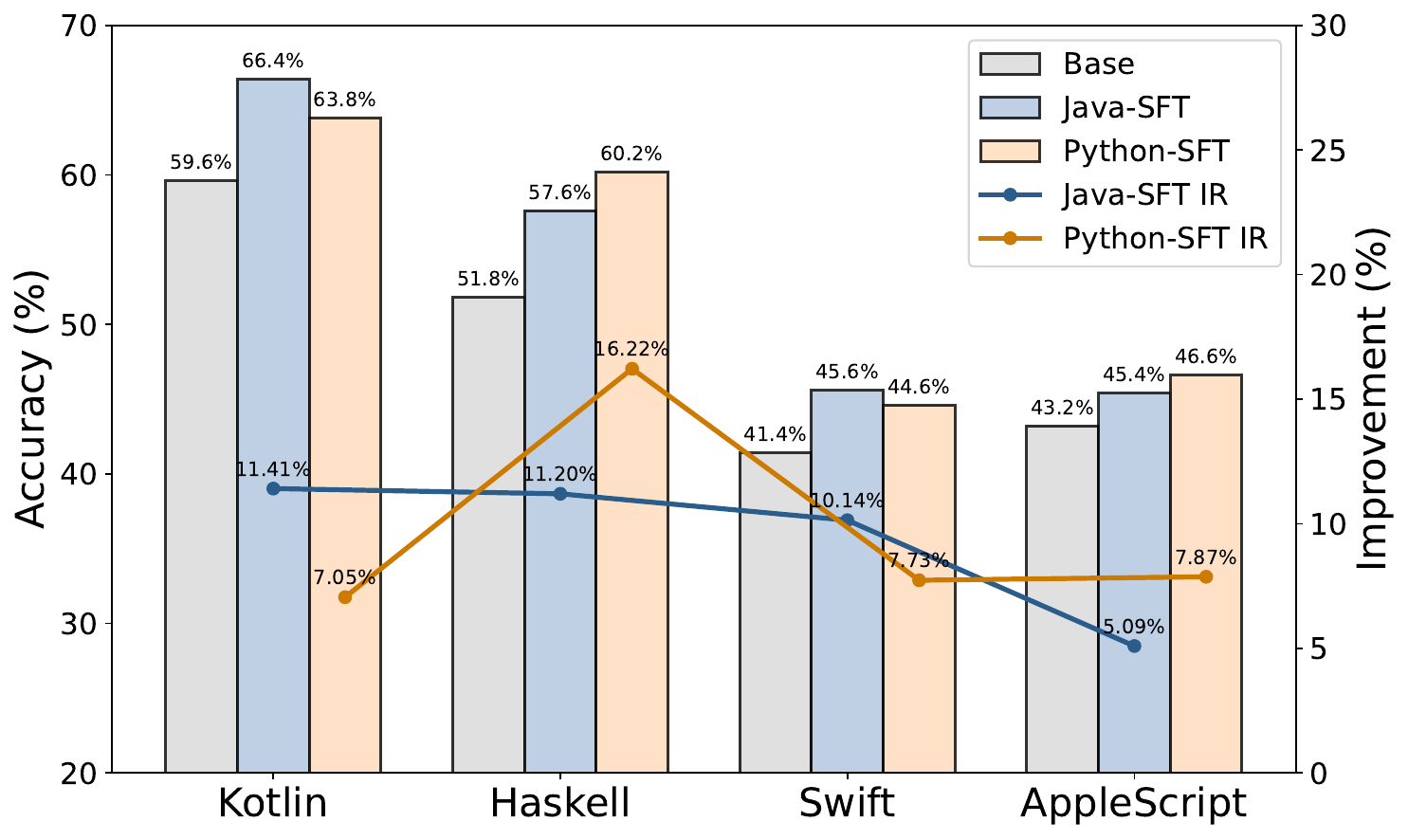}
    \caption{Code Search}
  \end{subfigure}
  \caption{Results of Transfer Learning for Low-resource Languages. \new{IR denotes the improvement rate, calculated as the relative performance gain over the non-fine-tuned baseline.}}
  \label{fig:transfer}
\end{figure}

\vspace{1.5mm}
\textbf{Results}.
As illustrated in Figure~\ref{fig:transfer}, fine-tuning StarCoderBase-15.5B with Java and Python data significantly improves the task performance across all evaluated low-resource languages. 
The Java-SFT model (fine-tuned with Java data) achieves performance improvements of 5.68\%$\sim$19.83\% in code summarization and 5.09\%$\sim$11.41\% in code search across Kotlin, Haskell, Swift, and AppleScript.  
The Python-SFT model (fine-tuned with Python data) similarly enhances performance, with gains of 7.05\%$\sim$16.22\% in code summarization and 4.89\%$\sim$17.61\% in code search.   
These results confirm successful knowledge transfer from high-resource languages (Java and Python) to their low-resource counterparts.

More significantly, our analysis reveals that linguistic proximity within the language family hierarchy strongly influences transfer effectiveness. In particular, Kotlin and Swift exhibit greater performance improvements with the Java-SFT model than the Python-SFT model in both tasks, consistent with their closer phylogenetic relationship to Java. Conversely, AppleScript demonstrates more substantial gains with the Python-SFT model, reflecting its stronger alignment with Python's language characteristics.



\new{To assess the robustness of these improvements, we perform additional statistical analyses across all evaluated languages. For each model configuration (Base, Java-SFT, Python-SFT), we execute three independent runs and compute the mean, standard deviation, and 95\% confidence interval of the BLEU and accuracy scores. Across the four target languages, standard deviations remain below 0.35 BLEU for summarization and below 0.9\% for search accuracy, indicating stable performance across runs. Paired t-tests comparing Java-SFT and Python-SFT against the base model reveal statistically significant differences (\(p < 0.05\)), with Cohen’s \(d\) values ranging from 0.42 to 0.71, corresponding to medium effect sizes. These findings confirm that the performance gains presented in Figure~\ref{fig:transfer} are consistent and unlikely to arise from random variation.}

\begin{tcolorbox}[enhanced, width=\linewidth, boxrule=0.8pt, 
 left=2pt, right=2pt, top=2pt, bottom=2pt, drop fuzzy shadow=black,]
\textbf{Finding 6:} The capability of code language models on low-resource languages can be substantially enhanced by fine-tuning on high-resource languages that are phylogenetically related.
\end{tcolorbox}
\vspace{2pt}

\subsection{Multilingual Curriculum Learning}
\label{sec:curriculum}

The uncovered programming language families provide practical guidance for optimizing fine-tuning sequences of code LLMs. Motivated by these insights, we propose a curriculum learning strategy tailored for multilingual code training. The central principle is that programming languages structurally closer to those already mastered (i.e., well-pretrained) by the model are easier to learn \cite{KWMR20, MauryaD22}. We therefore designate English as the base language and construct curricula that progressively transition from English to other languages according to their embedding-space proximity. By leveraging English-proficient base models, semantic similarity to English in the embedding space serves as a proxy for relative learning complexity and as a guiding signal for curriculum design.

To investigate the effect of training order, we design three comparative curricula:

\begin{enumerate}
    \item \textit{Near-to-Far}: Languages are introduced in descending similarity to English, thereby gradually increasing learning difficulty: AppleScript$\rightarrow$ Python$\rightarrow$ Swift$\rightarrow$ Kotlin$\rightarrow$ JavaScript$\rightarrow$ Go$\rightarrow$ Rust$\rightarrow$ Java$\rightarrow$ C++$\rightarrow$ Haskell.
    
    \item \textit{Far-to-Near}: Languages are introduced in ascending similarity to English, beginning with the most distant: Haskell$\rightarrow$ C++$\rightarrow$ Java$\rightarrow$ Rust$\rightarrow$ Go$\rightarrow$ JavaScript$\rightarrow$ Kotlin$\rightarrow$ Swift$\rightarrow$ Python$\rightarrow$ AppleScript.
    
    \item \textit{Random (Control)}: Languages are introduced in a random order without structured progression: 1) C++$\rightarrow$ Go$\rightarrow$ Rust$\rightarrow$ Haskell$\rightarrow$ Swift$\rightarrow$ AppleScript$\rightarrow$ Java$\rightarrow$ Python$\rightarrow$ Kotlin$\rightarrow$ JavaScript; \new{2) Python$\rightarrow$ Swift$\rightarrow$ C++$\rightarrow$ Java$\rightarrow$ Go$\rightarrow$ Kotlin$\rightarrow$ Haskell$\rightarrow$ AppleScript$\rightarrow$ Rust$\rightarrow$ JavaScript; 3) Rust$\rightarrow$ Kotlin$\rightarrow$ Haskell$\rightarrow$ C++$\rightarrow$ JavaScript$\rightarrow$ Go$\rightarrow$ Python$\rightarrow$ Swift$\rightarrow$ Java$\rightarrow$ AppleScript.} 
\end{enumerate}

We evaluate the effectiveness of these curricula through controlled experiments on representative code generation tasks.

\vspace{1.5mm}
\textbf{Experimental Setup.}
We employ an incremental fine-tuning approach for LLMs following the curriculum sequence. 
To ensure independent adaptation, we reset both the learning rate and optimizer states between language transitions.
Model performance is evaluated using CodeBLEU ~\cite{ren2020codebleumethodautomaticevaluation} for code generation tasks. Given our use of English as the foundational language, we employ LLaMA2-7B ~\cite{llama2} due to its proven proficiency in natural language processing. Our study encompasses ten programming languages: C++, Go, Rust, Haskell, Swift, AppleScript, Java, Python, Kotlin, and JavaScript.

To mitigate data contamination risks, we construct a carefully curated dataset from GitHub repositories, with all test samples postdating LLaMA2-7B's training cutoff. We collect file-processing code snippets and their corresponding comments for the aforementioned programming languages, with a total of 1,800 samples for each language. In each round of training and testing, we allocate 1,260 samples (70\%) for fine-tuning and reserve 540 samples (30\%) for evaluation.

\begin{figure}[t]
       \centering
       \includegraphics[width=0.6\textwidth, trim=0 0 0 0]{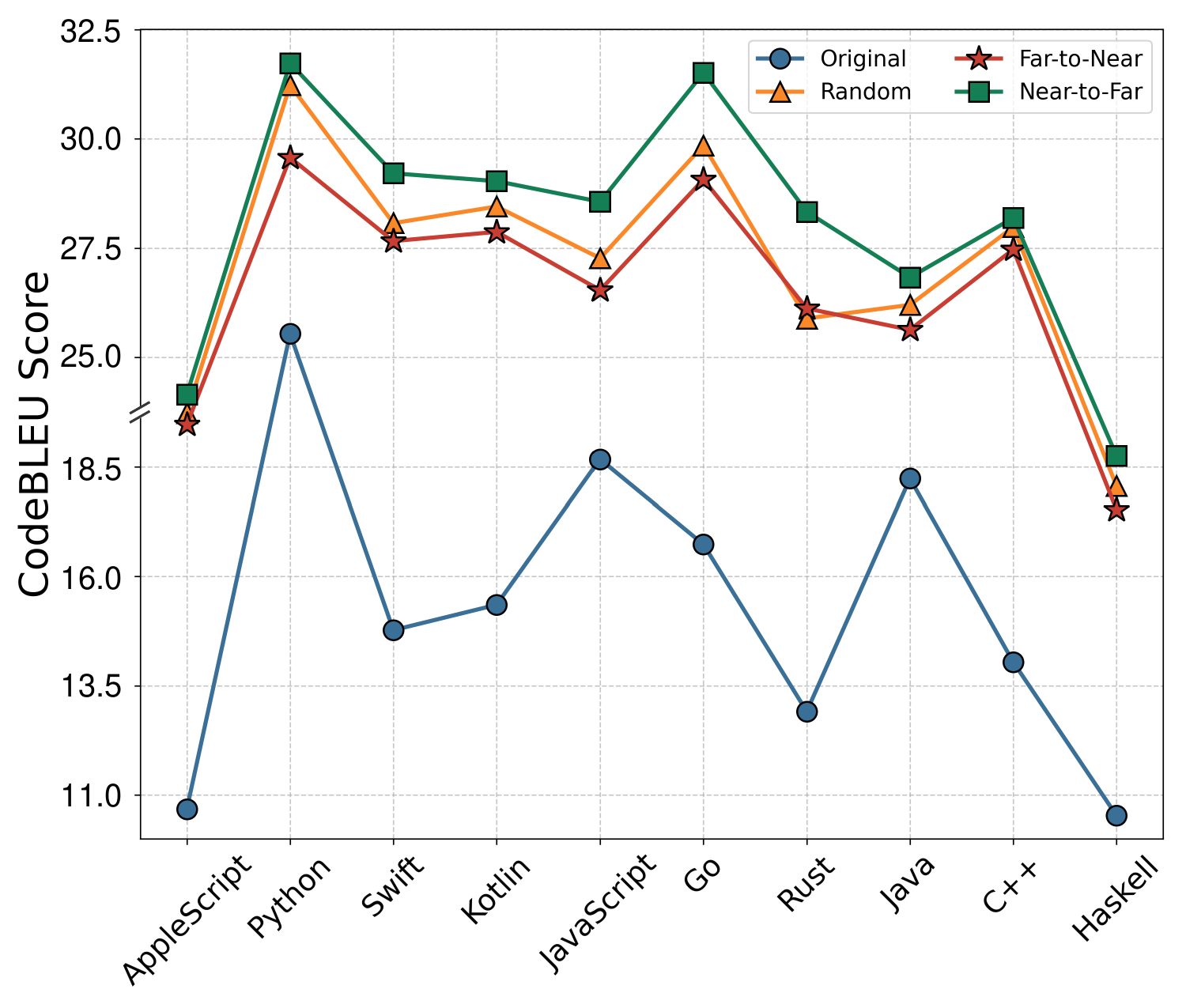}
       \caption{Results of Curriculum Learning for Code Generation.}        
       \label{fig:codeGeneration}
\end{figure}

\vspace{1.5mm}
\textbf{Results.}
As shown in Figure~\ref{fig:codeGeneration}, all curriculum-based fine-tuning strategies lead to substantial improvements in CodeBLEU scores compared to the original, non-finetuned model. Among them, the \textit{Near-to-Far} curriculum yields the best performance, achieving an average CodeBLEU score of 28.03 (82.67\% improvement over baseline), while consistently maintaining top-tier results across all evaluated programming languages.
\new{The \textit{Random} curricula rank second, with average CodeBLEU scores of 27.07, 27.25, and 26.90 (76.41\%, 77.58\%, and 75.30\% improvements), outperforming the \textit{Far-to-Near} curriculum, which achieves the lowest relative gain with an average score of 26.49 (a 72.63\% improvement). For clarity, only Random \#1 is included in the figure as a representative case.}

Across all fine-tuned models, Python and Go exhibit relatively strong performance. This phenomenon likely stems from two factors: (1) their natural language-like syntax reduces learning complexity for LLMs, and (2) their extensive adoption in practice provides richer training signals. Notably, Go exhibits substantial performance gains after fine-tuning, suggesting that its structural commonalities with other languages facilitate effective cross-lingual knowledge transfer during the curriculum learning process.

\begin{tcolorbox}[enhanced, width=\linewidth, boxrule=0.8pt, 
 left=2pt, right=2pt, top=2pt, bottom=2pt, drop fuzzy shadow=black,]
\textbf{Finding 7:} 
The linguistic similarity-based curriculum strategies significantly improve the performance of LLMs compared to alternative sequencing approaches.
\end{tcolorbox}
\vspace{2pt}

\subsection{Intermediary Code Translation}
\label{sec:translation}

Inspired by natural languages, where English often functions as an intermediary language for cross-linguistic communication, we aim to identify an optimal pivot programming language to facilitate LLM-mediated code translation. Such a pivot language could bridge the gaps between source and target programming languages, enabling smoother knowledge transfer between linguistically distant programming paradigms.

Specifically, for a given code snippet that requires translation, the LLM is first tasked with generating an intermediate translation in the selected pivot language. The intermediate output undergoes validation through a set of test cases to verify functional correctness. 
Once validated, the LLM proceeds to translate the intermediate representation into the target language, finalizing the translation process.

\vspace{1.5mm}
\textbf{Experimental Setup.}
We evaluate the effectiveness of different programming languages as pivot languages for LLM-based intermediary code translation. Our experiments focus on Python$\rightarrow$X$\rightarrow$Y translation tasks, where X serves as the pivot language and Y represents the target language.  
Python serves as the source language due to its established role in code intelligence benchmarks and the inherent challenges of direct Python$\rightarrow$Y translation~\cite{JiaoYLQGS23}.
The study investigates six pivot languages (C++, Go, Java, JavaScript, PHP, and Ruby), selected through a principled approach considering: their prevalence in production codebases ~\cite{CodeSearchNet}, representation of diverse programming paradigms, coverage of major type systems, and spanning both high- and low-resource languages. 

We employ Qwen2.5-Coder-14B~\cite{hui2024qwen2} as our base model, chosen for its demonstrated code generation capabilities and computational tractability.
The \emph{PolyHumanEval}~\cite{abs-2410-09812} benchmark is adopted for evaluation, which contains 164 carefully curated programming problems with language-specific test suites. The translation quality is quantified with Computational Accuracy (CA)~\cite{transcoder} metric, which is defined as the proportion of generated code samples that successfully pass all corresponding test cases.

\begin{table*}[!t]
    \centering
    \caption{Results of Intermediary Code Translations on Python$\rightarrow$Y Tasks}
    \label{tab:translation}
    \resizebox{\textwidth}{!}{
    \begin{tabular}{l l l l l l l l}
    \toprule
        \textbf{X} & \textbf{Py$\rightarrow$X$\rightarrow$C++} & \textbf{Py$\rightarrow$X$\rightarrow$Go} & \textbf{Py$\rightarrow$X$\rightarrow$Java} & \textbf{Py$\rightarrow$X$\rightarrow$JS} & \textbf{Py$\rightarrow$X$\rightarrow$PHP} & \textbf{Py$\rightarrow$X$\rightarrow$Ruby} & \textbf{Avg.} \\
        \midrule
        None & 87.80 & 87.80 & 80.49 & 87.80 & 85.98 & 89.02 & 86.48 \\
C++    &   --  & \textbf{89.63(+1.83)} & 90.24(+9.75) & 90.24(+2.44) & \textbf{90.85(+4.87)} & 87.80(-1.22) & 89.75(+3.53) \\
Go     & \textbf{88.41(+0.61)} &   --  & \textbf{90.85(+10.36)} & 89.63(+1.83) & 89.63(+3.65) & 92.68(+3.66) & \textbf{90.24(+4.02)} \\
Java   & 87.80(+0.00) & 88.41(+0.61) &   --  & \textbf{92.07(+4.27)} & 90.24(+4.26) & \textbf{96.95(+7.93)} & 91.09(+3.41) \\
JS     & 87.80(+0.00) & 85.98(-1.82) & 75.00(-5.49) &  --   & 84.15(-1.83) & 91.46(+2.44) & 84.88(-1.34) \\
PHP    & 85.98(-1.82) & 80.49(-7.31) & 89.02(+8.53) & 86.59(-1.21) &   --  & 88.41(-0.61) & 86.10(-0.48) \\
Ruby   & 85.98(-1.82) & 85.37(-2.43) & 85.37(+4.88) & 89.02(+1.22) & 85.37(-0.61) &   --  & 86.22(+0.25) \\
    \bottomrule
    \end{tabular}
    }
\end{table*}

\vspace{1.5mm}
\textbf{Results.}
Table \ref{tab:translation} summarizes the experimental results. It reveals significant variations in the efficacy of different programming languages as intermediate representations.
Go and Java demonstrate superior performance as pivot languages, achieving computational accuracy improvements of at least 3.4 percentage points compared to the baseline. Among these, Go yields the most substantial enhancement (4.02 points), consistently improving translation quality across all target languages. 
\new{The primary rationale is that Go's observed advantage aligns closely with its position in the learned embedding space (Figure~\ref{fig:matrix}), where it occupies a central and structurally neutral region between imperative and functional paradigms and exhibits the highest average similarity (0.39) to other languages. This cross‑lingual proximity enables Go to preserve semantic and type‑level information more faithfully during intermediate translation, thereby facilitating more stable knowledge transfer. The consistent improvements observed across multiple translation directions further confirm Go’s role as a versatile pivot language, rather than a task‑specific artifact.}

In contrast, JavaScript, PHP, and Ruby yield performance comparable to or even worse than direct translation when used as pivot languages. For PHP and Ruby, the relatively low average similarity with the target languages limits their effectiveness in improving translation quality. Although JavaScript exhibits higher similarity to many target languages, its performance as an intermediary remains unsatisfactory. Error analysis reveals that most failures in translations involving JavaScript are caused by type mismatches. This may be due to JavaScript’s dynamic and weak typing, which makes it difficult to convey accurate type information during translation.

\begin{tcolorbox}[enhanced, width=\linewidth, boxrule=0.8pt, 
 left=2pt, right=2pt, top=2pt, bottom=2pt, drop fuzzy shadow=black,]
\textbf{Finding 8:} 
The efficacy of intermediary code translation exhibits a strong correlation with the linguistic proximity between pivot and target languages. \textit{Go} emerges as the optimal universal intermediary due to its central position in the programming language taxonomy.
\end{tcolorbox}
\vspace{2pt}

\section{Discussion}

\subsection{Future Directions}
The implications of programming language families lie in the wider field of both code LLMs and programming languages.
In this section, we delineate several future directions that could help advance the field.

\subsubsection{Intra-family and Inter-family Learning}

Greater emphasis could be placed on in-family and cross-family learning. Intra-family learning aims to develop unified code LLMs for comprehension, generation, and translation across languages within the same families, leveraging shared linguistic features to enhance low-resource language support. 
Inter-family learning, on the other hand, seeks to train code LLMs to capture the diversity of programming languages by selecting representative languages from each family, optimizing adaptability while minimizing training resource requirements.

\subsubsection{Design AI-friendly Language}

Future research could explore the development of novel programming languages specifically optimized for compatibility with LLMs, referred to as AI-friendly languages. Interdisciplinary collaboration with cognitive science would further align language design with both human cognition and LLM inference mechanisms. This direction seeks to improve the efficiency and effectiveness of human-AI pair programming.

\subsubsection{Programming Linguistics Study} 

While the data-driven approach accurately models semantic relationships in practical usage, it does not always align with historical, paradigmatic, or type-theoretic classifications. A promising research direction involves synthesizing these complementary perspectives to develop a robust programming language taxonomy. Such integration would simultaneously advance syntactic and intrinsic language analysis while facilitating the exploration of interrelationships and shared characteristics across these linguistic domains.

\subsection{Limitations and Threat to Validity}

This study has several limitations that could affect the validity of our findings: Firstly, we derive programming language embeddings from code snippets crafted using 21 meticulously designed features. While these linguistic features effectively capture fundamental syntactic elements of programming languages, they may not completely characterize all language-specific properties. We reserve the refinement of this feature set to more accurately model both shared patterns and unique aspects of programming languages for future research.

Secondly, our study is restricted to 19 programming languages. While they encompass most mainstream languages and programming paradigms, they may benefit from broader language inclusion to enhance the generalizability of our insights. 
Additionally, we encode programming language representations using GPT-4o and utilize StarCoderBase-15.5B, LLaMA2-7B, and Qwen2.5-Coder-14B for four code-intelligent tasks; however, insights derived from these models and tasks may not fully generalize to others. 
While we believe our findings are generalizable, future work should explore a wider variety of languages, tasks, and models to strengthen validity.

Thirdly, due to the absence of an established ``ground truth'' for programming language families, it is challenging to objectively assess the accuracy and quality of our work. To mitigate this threat, we conduct our validation from two perspectives: 1) by demonstrating its alignment with both conventional programming language classifications and findings from prior model neuron analysis, and 2) by illustrating how the inferred relationships among languages can inform language selection and training strategies in multi-lingual code intelligence tasks.

\section{Conclusion}
This paper presents a novel embedding-based framework for quantifying programming language relationships using parallel code representations across 21 linguistic features. To the best of our knowledge, this represents the first comprehensive study of programming language taxonomy through automated AI techniques. Our method uncovers six language families from 19 programming languages, with \textit{Go} exhibiting the highest centrality. This taxonomy enables optimized language selection and curriculum design for multilingual code LLM training and inference. Experiment results on four tasks (\emph{i.e.,} code summarization, code search, code generation, and code translation) demonstrate consistent performance improvements. 
We hope this study can inspire future research along this direction.

\section*{Data Availability}
All the code and data used in our study are publicly available at: 
\url{https://github.com/magicalYY/artifact-fse26}. It includes (i) the code for generating code samples, computing embeddings, and performing clustering analysis; (ii) the code samples generated based on 21 linguistic features across 19 programming languages; and (iii) all experimental datasets and results.

\begin{acks}
This research is funded by the National Key Research and Development Program of China (Grant No. 2023YFB4503802), the National Natural Science Foundation of China (Grant No. 62232003), and the Natural Science Foundation of Shanghai (Grant No. 25ZR1401175).
\end{acks}

\bibliographystyle{ACM-Reference-Format}
\bibliography{ref}

\end{document}